\documentclass[twocolumn,showpacs,preprintnumbers,amsmath,amssymb,aps]{revtex4}
\usepackage{graphicx}
\usepackage{dcolumn}
\usepackage{bm}

\begin{document}

\title{Comparative study of defect energetics in HfO$_2$ and SiO$_2$}

\author{W. L. Scopel}
\author{Ant\^onio J. R. da Silva}
\author{W. Orellana}
\author{A. Fazzio}
\email{fazzio@if.usp.br}
\date{\today}

\begin{abstract}

We perform \textit{ab initio} calculations, based on density functional
theory, of substitutional and vacancy defects in the monoclinic hafnium
oxide ({\it m}-HfO$_2$) and $\alpha$-quartz (SiO$_2$). The neutral oxygen vacancies and
substitutional Si and Hf defects in HfO$_2$ and SiO$_2$, respectively,
are investigated. Our calculations show that, for a large range of Hf
chemical potential, Si substitutional defects are most likely to form
in HfO$_2$, leading to the formation of a silicate layer at the
HfO$_2$/Si interface. We also find that it is energetically more favorable
to form oxygen vacancies in SiO$_2$ than in HfO$_2$, which implies that
oxygen deficient HfO$_2$ grown on top of SiO$_2$ will consume oxygen
from the SiO$_2$.

\end{abstract}

\maketitle
The continuous device miniaturization in the microelectronic industry
will eventually lead, within the present technology, to the end of the use of
amorphous SiO$_2$ ({\it a}-SiO$_2$) as gate dielectric in
metal-oxide-semiconductor field-effect transistors (MOSFETs).
The existence of a thickness limit for the
{\it a}-SiO$_2$ around 10-12 \AA, has clearly been established
experimentally \cite{Muller}. One way to circumvent this problem, still
keeping Si as the basic
device material, is to employ high-permitivity materials as alternative gate
dielectrics in place of the conventional {\it a}-SiO$_2$. Among them,
hafnium oxide
is emerging as the material with greatest potential to substitute SiO$_2$,
mainly due to its high dieletric constant
and thermodynamic stability, when it forms interface with Si.

Even though hafnium oxide is thermodynamically stable against an overall
decomposition as Hf and SiO$_2$ when grown on Si, interfacial reactions can occur.
Thereby the formation of a thin interfacial layer (oxides, silicates and silicides)
between the HfO$_2$ and the Si surface, has been recently observed \cite{Roh,Jeon}.
This interfacial layer occurs during almost any film growth processes or
post-annealing, which is an intrinsic part of any growth cycle. Therefore, the
thermodynamic stability of the hafnium oxide in contact with silicon is identified
as a critical issue for the application of alternative gate dielectric in
silicon-based devices \cite{Misa,Bush}. Moreover, the study of possible defects
related to the migration of atoms across the interface is of fundamental
importance. In particular, a significant source of defects in this system is
the interface itself, which has been shown \cite{Gutowski} to consist of Hf
silicates with dielectric constant lower than that of HfO$_2$ \cite{Callegari,Kato}.

In the present work we address the formation of neutral defects through
first-principles calculations, based on the density functional theory (DFT).
We analyze the formation of Si substitutional defects in HfO$_2$, as well as
Hf substitutional defects in SiO$_2$, for different growth conditions. Finally,
the energetics of an oxygen vacancy in SiO$_2$ is compared to a similar vacancy
in HfO$_2$, in order to understand the growth of hafnium oxide under oxygen-poor
conditions.

Many experimental works \cite{Cho,Park,Moonju,Bastos} have addressed
the chemical reactions that could occur in the HfO$_2$/Si
interface during the HfO$_2$ growth cycle. This is due in order to
prevent and/or control the interfacial layer formation. Almost all
works have reported the formation of an interfacial Hf silicate in
oxygen-rich atmospheres. Furthermore,
Wang and co-workers \cite{Lim} have shown that in opposite conditions, {\it i.e.},
oxygen-deficient atmospheres, the Hf silicate interfacial formation does not happen
during the HfO$_2$ growth cycle. However, there are still many open questions,
such as which atomic species are migrating when the interfacial silicate is formed.
For example, it is important to know if Hf will be incorporated in a formed
SiO$_2$ layer, or if Si from either the bulk Si or this SiO$_2$ layer will be
incorporated in the HfO$_2$.

The DFT calculations were performed using ultrasoft Vanderbilt
pseudopotentials \cite{Vanderbilt}, and the generalized gradient
approximation(GGA) for the exchange-correlation potential as implemented
in VASP code \cite{Perdew,vasp1,vasp2,vasp3}. In order to study the defects
in the different systems we have considered the monoclinic HfO$_{2}$ and
the $\alpha$-quartz (SiO$_2$) crystalline phases, using a 96 atoms and 72
atoms supercells, respectively. For these cells, we have used a plane wave
cutoff energy of 400 eV and a 2$\times$2$\times$2 Monkhorst-Pack {\bf{k}}-mesh.
These crystal structures have been previously used to describe these
systems \cite{nosso1,nosso2,foster}. In all calculations the atoms were allowed
to relax until all components of the atomic forces were smaller than
0.025 eV/\AA.

The silicon substitutional defect (Si$_{\text{Hf}}$) was created in the
\textit{m}-HfO$_{2}$ supercell by substitution of one hafnium atom by
one silicon atom in the equilibrium perfect crystal. On the other hand, the
hafnium substitutional defect (Hf$_{\text{Si}}$) was created in the
$\alpha$-quartz by substitution of one silicon atom by one hafnium atom
in the equilibrium perfect crystal. Our results show that the presence of
either Si$_{\text{Hf}}$ or Hf$_{\text{Si}}$ in HfO$_2$ or SiO$_2$,
respectively, do not introduce any additional active levels in the band gap.
For the perfect HfO$_2$ at the equilibrium, we obtain an indirect band
gap of 3.9 eV along $\Gamma$-B. For this same supercell, the density of
states (DOS) shows that the O(\textit{2s}) and O(\textit{2p}) bands are
centered at around -14 and 0 eV, with bandwidths of approximately
4 eV and 8 eV, respectively, below valence-band edge, whereas the
Hf(\textit{5d}) band is centered at around 8 eV forms in the conduction
band, with bandwidth of 6 eV. These results
are in good agreement with previous DFT calculations of this
material \cite{kang}. Moreover, for HfO$_2$ with a Si substitutional
defect, we observe a resonant state around 6.9 eV below the top of
the valence band, which is related to the O(\textit{2p})-like and \textit{Si}
bonding state \cite{Binggeli}.

The formation energy for a Si$_{\text{Hf}}$ is calculated as
\begin{eqnarray}
E_{f}^{Si_{\text{Hf}}} = [E_{t}(Si_{\text{Hf}})+\mu_{Hf}]-[E_{t}(HfO_2)+\mu_{Si}],
\label{E_formation}
\end{eqnarray}
\noindent whereas for Hf$_{\text{Si}}$ the similar expression is
\begin{eqnarray}
E_{f}^{Hf_{\text{Si}}} = [E_{t}(Hf_{\text{Si}})+\mu_{Si}]-[E_{t}(SiO_2)+\mu_{Hf}].
\label{E_formation2}
\end{eqnarray}
In the above expressions, E$_{t}$($D_S$) are the total energies of the fully
relaxed supercells (either \textit{m}-HfO$_2$ or SiO$_2$) with the
substitutional defect $D_S$, and E$_{t}$($XO_2$) are the total energies of the
similar supercells for the perfect crystals of XO$_2$ (X = Hf or Si). The
values of the chemical potentials, $\mu_{Hf}$ and $\mu_{Si}$, depend on the
growth conditions. We have considered two limits for $\mu_{Hf}$: (i) The
bulk metal as a reference, which would correspond to a Hf-rich growth condition
and the formation of Hf clusters in the bulk or at the surface of the oxide;
(ii) Under oxygen rich conditions, or considering that there is plenty of
oxygen atoms in the HfO$_2$ \cite{nieminen}, such that the removed hafnium
remains always in equilibrium with the oxygen gaseous, the $\mu_{Hf}$(gas) can be
obtained as $\mu_{Hf}$(gas) = $\mu_{HfO_{2}} - \mu_{O_{2}}$. The HfO$_2$ chemical
potential ($\mu_{HfO_{2}}$) was obtained as the energy per unit formula
for the monoclinic bulk hafnium oxide,
and $\mu_{O_{2}}$ is the energy of an isolated oxygen molecule, which was
obtained through a DFT total energy calculation for an O$_2$ inside a
cubic supercell of 15\,\AA\ side. For the silicon chemical potential, two
similar limits were considered: (i) $\mu_{Si}$ as the crystalline bulk Si
chemical potential, and (ii) $\mu_{Si}$ as the chemical potential
of Si in SiO$_2$ under oxygen rich conditions, {\it i.e.},
$\mu_{Si}$ = $\mu_{SiO_{2}} - \mu_{O_{2}}$, with $\mu_{O_{2}}$ as above and
$\mu_{SiO_{2}}$ obtained as the energy per unit formula of $\alpha$-quartz.

\begin{figure}[h]
\includegraphics[width=8.5cm]
{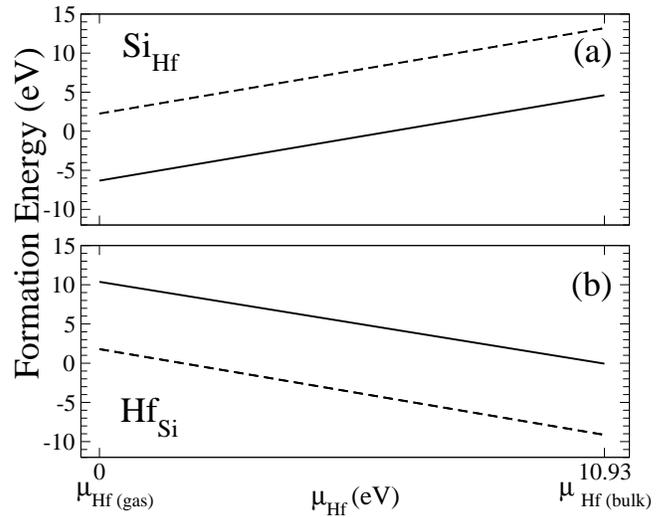}
\caption{\label{chemical_pot} Formation energies for the two substitutional
defects considered in this work: (a) Si in place of a Hf in HfO$_2$, and
(b) Hf in place of a Si in SiO$_2$. The formation energies are plotted as a
function of the Hf chemical potential, and for two values of the Si chemical
potential, the bulk Si chemical potential (solid curves) and the chemical
potential for Si in the SiO$_2$ under an oxygen rich environment
(dashed curves).}
\end{figure}

The results for the formation energies are presented in Fig.~\ref{chemical_pot}.
As can be seen from Fig.~\ref{chemical_pot}(a), under oxygen rich conditions, the
formation of a Si substitutional defect in HfO$_2$ is very likely to occur,
specially if the Si source is the bulk crystal. Under this O-rich situation,
the formation of this defect would only become unfavorable if a thick enough
layer of SiO$_2$ would exists between the Si substrate and the HfO$_2$. In this
Case, either the source of Si atoms would be the silica, or the Si would
somehow not diffuse through this thick layer. If the SiO$_2$ is thin enough,
such that the Si diffusion through it is significant, then the formation of a
Hf silicate (at least close to the interface) seems unavoidable \cite{lee}.
On the other hand, under Hf rich conditions, it is less likely that Si
substitutional defects in HfO$_2$ will be formed.

The formation of Hf substitutional defects in SiO$_2$ has a behavior that is
opposite to Si$_{\text{Hf}}$, as can be seen in Fig. 1(b), {\it i.e.}, it is not
likely to form under oxygen rich conditions, and becomes more probable under
Hf-rich conditions. In a situation where the Si chemical potential is given
by its bulk value (either thin or no SiO$_2$ layer), the formation of
a Hf$_{\text{Si}}$ in SiO$_2$ is unlikely to happen, for a large range of Hf
chemical potential.

We also considered the formation of neutral oxygen vacancies ($V_O$),
both in SiO$_2$ as well as in HfO$_2$ \cite{foster}, since they can be
created in films and bulk samples due to the growth cycle. The neutral
oxygen vacancy in the {\it m}-HfO$_{2}$ and $\alpha$-quartz were
generated by simple removal of an oxygen atom, followed by full
relaxation of all remaining atoms. The formation energies for a
$V_O$ in XO$_2$ (X=Hf or Si), $E_f(V_O)$, were calculated as,
\begin{eqnarray}
E_{f}^{XO_2}(V_O)= [E_t^{XO_2}(V_O)+\mu_{O}]-[E_{t}(XO_2)],
\label{E_vacancy}
\end{eqnarray}
\noindent where $E_{t}^{XO_2}(V_O)$ and $E_{t}(XO_2)$, are the total energies
of supercells of $XO_2$ (X = Hf or Si) with, and without, an oxygen vacancy,
respectively. The oxygen chemical potential, $\mu_{O}$, was considered either
as the total energy of an isolated oxygen atom, or as one half of the energy
of an isolated oxygen molecule (in both cases, the O$_2$ triplet ground state was
used). The monoclinic phase has non-equivalent oxygen atoms, \textit{i.e}, which
some sites are threefold coordinated, whereas others are fourfold coordinated by
hafnium atoms. In this way, we have determined the formation energy for
both vacancies types using Eq.~(\ref{E_vacancy}) and we obtain a formation
energy difference around 0.02 eV.
In the former case, we obtained of $E_f^{HfO_2}(V_O)$ = 9.32 eV and
$E_f^{SiO_2}(V_O)$ = 8.10 eV, whereas for the latter choice of $\mu_{O}$, we
obtained $E_f^{HfO_2}(V_O)$ = 6.38 eV and $E_f^{SiO_2}(V_O)$ = 5.16 eV. This
indicates that, although neutral oxygen vacancies are energetically unfavorable in
both materials, they are more stable in silicon oxide than in hafnium oxide,
by approximately 1.23 eV. This implies that, if an oxygen deficient HfO$_2$
is grown on top of a SiO$_2$ layer, oxygen atoms will migrate from SiO$_2$.
In this way, oxygen vacancies are created in the silica layer toward
the HfO$_2$, healing at the sometimes oxygen vacancies in the hafnia.
Indications that this process does indeed occur, has been recently
reported \cite{Lim}. 

In summary, our results show that, unless the hafnium chemical potential is
always very close to its bulk value,{\it i.e.}, oxygen-poor growth conditions,
the formation of Si substitutional defects in the HfO$_2$ is almost unavoidable.
This will lead to the formation of a silicate-like layer close to the
Interface. For a very large range of Hf
chemical potential it is very likely that incoming silicon atoms from the Si
substrate will form a (HfO$_2$)$_x$(SiO$_2$)$_y$ layer, as supported by
experimental observations \cite{lee}. The only way to prevent the formation
of a silicate layer would be through the use of an oxygen deficient HfO$_2$,
as has been observed \cite{Lim}. However, even if this silicate formation
is prevented in the initial steps of dielectric growth, it seems almost impossible that it will not be formed given the necessary further
thermal processing steps, unless a barrier for the Si diffusion being
introduced \cite{Bastos}.

\begin{acknowledgments}

This research was supported by Brazilian agencies FAPESP and CNPq. We thank
CENAPAD-SP for computer time.

\end{acknowledgments}


\newpage

\thebibliography{article}

\bibitem{Muller}
D. A. Muller, T. Sorsch, S. Moccio, H. F. Baumann,
K. Evans-Lutterodt, and G. Timp, Nature {\bf 399}, 758 (1999).

\bibitem{Roh}
M.-H. Cho, Y. S. Roh, C. N. Whang, K. Jeong, S. W. Nahm,
D.-H. Ko, J. H. Lee, N. I. Lee and K. Fujihara, Appl. Phys. Lett.
{\bf 81}, 472 (2002).

\bibitem{Jeon}
T. S. Jeon, J. M. White, and D. L. Kwong, Appl. Phys. Lett. {\bf 78},
368 (2001).

\bibitem{Misa}
V. Misa, G. Lucovsky,and G. Parsons, MRS Bull. {\bf 27}, 212 (2002).

\bibitem{Bush}
B. W. Bush, O. Pluchery, Y. J. Chabal, D. A. Muller, R. L. Opila,
J. R. Kwo, and E. Garfunkel, MRS Bull. {\bf 27}, 206 (2002).

\bibitem{Gutowski}
M. Gutowski, J. E. Jaffe, C.-L. Liu, M. Stoker, R. I. Hegde,
R. S. Raj, and P. J. Tobin, Appl. Phys. Lett. {\bf 80}, 1897 (2002).

\bibitem{Callegari}
A. Callegari, E. Cartier, M. Gribelyuk, H. F. Okorn-Schmidt, and T. Zabel,
J. Appl. Phys. {\bf 90}, 6466 (2001). 

\bibitem{Kato}
H. Kato, T. Nango, T. Miyagwa, T. Katagiri, K. Soo Seol, and Y. Ohki,
J. Appl. Phys. {\bf 92}, 1106 (2002).

\bibitem{Bastos}
K. P. Bastos, J. Morais, L. Miotti, R. P. Pezzi, G. V. Soares,
I. J. R. Baumvol, R. I. Hegde, H. H. Tseng, and P. J. Tobin,
Appl. Phys. Lett. {\bf 81}, 1669 (2002).

\bibitem{Cho}
B. K. Park, J. Park, M. Cho, C. S. Hwang, K. Oh, Y. Han, and D.Y. Yang,
J. Electrochem. Soc. {\bf 149}, G89 (2002). 

\bibitem{Park}
B. K. Park, J. Park, M. Cho, C. S. Hwang, K. Oh, Y. Han, and D. Y. Yang,
Appl. Phys. Lett. {\bf 80}, 2368 (2002). 

\bibitem{Moonju}
M. Cho, J. Park, H. K. Park, and C. S. Hwang, Appl. Phys. Lett.
{\bf 81}, 3630 (2002). 

\bibitem{Lim}
S. J. Wang, P. C. Lim, A. C. H. Huan, C. L. Liu, J. W. Chai,
S. Y. Chow, J. S. Pan, Q. Li, and C. K. Ong, {\bf 82}, 2047 (2003).

\bibitem{Vanderbilt}
D. Vanderbilt, Phys. Rev. B {\bf 41}, 7892 (1990).

\bibitem{vasp1}
G. Kresse and J. Hafner, Phys. Rev. B {\bf 47}, 558 (1993). 

\bibitem{vasp2}
G. Kresse and J. Hafner, Phys. Rev. B {\bf 48}, 13115 (1993).

\bibitem{vasp3}
G. Kresse and J. Furthmuller, Comput. Mater. Sci. {\bf 6}, 15 (1996). 

\bibitem{Perdew}
J. P. Perdew, J. A. Chevary, S. H. Vosko, K. A. Jackson, M. R. Pederson,
D. J. Singh, and C. Fiolhais, Phys. Rev. B {\bf 41}, 6671 (1992).

\bibitem{nosso1}
W. Orellana, A. J. R. da Silva, and A. Fazzio, Phys. Rev. Lett.
{\bf 87}, 155901 (2001).

\bibitem{nosso2}
W. Orellana, A. J. R. da Silva, and A. Fazzio, Phys. Rev. Lett.
{\bf 90}, 016103 (2003).

\bibitem{foster}
A. S. Foster, F. L. Gejo, A. L. Shluger, and R. M. Nieminen,
Phys. Rev. B {\bf 65}, 174117 (2002).

\bibitem{kang}
Kang, J., Lee, E.-C., and Chang, K.J., Phys. Rev. B {\bf 68},
054106 (2003).

\bibitem{Binggeli}
N. Binggeli, N. Troullier, J. L. Martins, and J. Chelikowsky,
Phys. Rev. B {\bf 44}, 4771 (1991). 

\bibitem{nieminen}
A.S. Foster, A.L. Shluger, and R.M. Nieminen, Phys. Rev. Lett.
{\bf 89}, 225901 (2002).

\bibitem{lee}
J.-H. Lee, N. Miyata, M. Kundu, and M. Ichikawa, Phys. Rev. B {\bf 66},
233309 (2002).
\end{document}